\documentclass[review,12pt]{elsarticle}

\usepackage{lineno,hyperref}
\modulolinenumbers[5]

\usepackage{graphicx}
\usepackage{bm}
\usepackage{verbatim}
\usepackage{amsmath}
\usepackage{amssymb}
\hypersetup{colorlinks=false}

\journal{Journal of Magnetism and Magnetic Materials}

\begin{document}

\begin{frontmatter}

\title{The random field Blume-Capel model revisited}


\author{P. V. Santos \fnref{myfootnote} \corref{mycorrespondingauthor}}
\cortext[mycorrespondingauthor]{Corresponding author}
\ead{priscilavs65@yahoo.com.br}
\fntext[myfootnote]{Centro de Forma\c{c}\~{a}o de Professores, Universidade Federal do Rec\^{o}ncavo da Bahia, Amargosa- BA, 45300-000, Brazil}

\author{F. A. da Costa}
\ead{fcosta@dfte.ufrn.br}

\author{J. M. de Ara\'{u}jo }
\ead{joaomedeiros@dfte.ufrn.br}

\address{Departamento de F\'{i}sica Te\'{o}rica e Experimental \\ Universidade Federal do Rio Grande do Norte \\ Natal-RN, 59078-900, Brazil}

\begin{abstract}
We have revisited the mean-field treatment for the Blume-Capel model under the presence of a discrete random magnetic field as introduced by Kaufman and Kanner \cite{Kaufman}.
The magnetic field ($H$) versus temperature ($T$) phase diagrams for given values of the crystal field $D$ were recovered in accordance to Kaufman and Kanner original work.
However, our main goal in the present work was to investigate the distinct structures of the crystal field versus temperature phase diagrams as the random magnetic field is varied because similar models have presented reentrant phenomenon due to randomness.
Following previous works we have classified the distinct phase diagrams according to five different topologies. The topological structure of the phase diagrams is maintained for both $H-T$ and $D-T$ cases.
Although the phase diagrams exhibit a richness of multicritical phenomena we did not found any reentrant effect as have been seen in similar models.

\end{abstract}



\end{frontmatter}

%
%
\newpage
\section{Introduction}
Disordered magnetic systems represent a great challenge in condensed matter physics since their properties are richer and more complex than their pure, non-disordered, counterparts \cite{DeDominicis_Giardina}. In particular, multicritical behavior and reentrance phenomena in disordered magnetic systems have been the subject of recent studies \cite{Gulpinaretal, Acharyya, daSilva, Sumedha, Carvalho, Costabile, Santosetal}. Due to both theoretical and experimental importance some attention has been devoted to models under the presence of random fields \cite{DeDominicis_Giardina,Nishimori,Kaufman2}. Most of these studies have considered the Blume-Capel (BC) model, proposed independently by Blume \cite{Blume} and Capel \cite{Capel}. In the pure ferromagnetic case, the BC model is an extension of the Ising model for spin-1 which takes into account the effect of a crystal field.  Its phase diagram presents both continuous and first-order transitions lines merging at a tricritical point \cite{BlumeEmeryGriffiths}. The experimental interest in this model has increased as well and we can mention several studies on systems as metallic alloys \cite{Danielle, Lara09, Blatter, Cowlam}, magnetic thin films \cite{Mazzitello}, metamagnets as $Ni(NO_3)_2$ \cite{BlumeEmeryGriffiths}, superconducting films \cite{Goldman}, dysprosium aluminium garnet \cite{Giordano}, liquid crystals \cite{Johari, Taylor} and others. The Blume-Capel model has been studied by several different methods including mean-field theory \cite{Blume, Capel}, renormalization group calculations (\cite{Antenucci} and references therein) and Monte Carlo simulations \cite{Mahan, Clusel}. An interesting theoretical question is how the phase diagrams are affected by the presence of quenched disorder \cite{Berker, Hui, Imry}. Several authors have considered the effect of random crystal field adopting, for this purpose, different approaches as well as different choices of the random field distribution \cite{ Lara12, Branco97, Benyoussef87, Carneiro89, Salmon10, Boccara, Maritan, Kaneyoshi, Buzano, Borelli}. In general, under the presence of quenched randomness the phase diagrams present a rich behavior with both continuous and coexistence lines, multicritical points and, in some cases, reentrance effects. Recently, Salmon and Tapia \cite{Salmon10} have studied the multicritical behavior in an infinite-range version of the BC model with the inclusion of quenched disorder in the crystal field and presented a classification of the possible phase diagrams according to their topology. 
Some recent studies have considered the Blume-Capel model in random fields \cite{Albayrak11, Kaya, Zhang}. 
As far as we know,  the Blume-Capel model under a bimodal $\pm H$ random
field were first obtained by Kaufman and Kanner  \cite{Kaufman}. They obtained several
$H-T$ phase diagrams for different values of the crystal field. Their phase diagrams do not 
present any reentrance effect. Thus, we decided to revisit the Kaufman and Kanner
results in order to investigate the corresponding $D$ (representing the uniform crystal field)
versus $T$ phase diagrams for different values of $H$. For this purpose, we 
consider an infinite-range version of the Blume-Capel model under a bimodal random
field. The resulting free energy density obtained by Kaufman and Kanner is recovered
in the thermodynamic limit. In addition to reproducing those results, we have also obtained $D-T$
phase diagrams for given values of $H$. Differently from what has been observed in other disordered spin-1 models \cite{Santosetal,daCosta,Ferrari}, no reentrant behavior was found in the present case. The outline of this work is organized as follows: In Section 2 we introduce Blume-Capel model under a random magnetic field and present the basic equations. In Section 3 we present our findings for the $D-T$ phase diagrams. Finally, we summarize our conclusions in Section 4.

%
%
\section{The model and basic equations}
Let us consider the Blume-Capel model described by the Hamiltonian \cite{Kaufman}
\begin{equation}
\mathcal{H}=-\frac{J}{2N}\sum_{(i,j)} S_{i}S_{j}+JD\sum_{i}S^{2}_{i}-J\sum_{i}H_{i}S_{i},
\label{eq:hamiltonian_kk}
\end{equation}
\noindent
where the spin variables $S_i$ assume the values -1, 0 and +1 at each site,
$(i,j)$ represents a sum over all distinct pairs of sites, $J > 0$ is the ferromagnetic exchange interaction (the factor $1/N$  is to ensure the proper thermodynamic limit).
The local fields $H_i$ are quenched, independent and identically distributed random variables which obey the bimodal distribution:
\begin{equation}
P(H_i)=\frac{1}{2}[\delta (H_{i}-H)+ \delta (H_{i}+H)].
\label{eq:bimodal}
\end{equation}
The free energy can be computed via the standard replica method \cite{Binder and Young}.  
In the thermodynamic limit the free energy per spin, in unities of $J$, becomes

\begin{equation}
\begin{split}
f =\frac{1}{2}m^2 &-\frac{1}{2}t \ln\left\{1+ 2\textrm{e}^{-d/t}\cosh\left[(m+h)/t\right]\right\} \\ & -\frac{1}{2}t \ln \left\{1+2\textrm{e}^{-d/t}\cosh\left[(m-h)/t\right]\right\}.
\end{split}
\label{eq:lim_termo}
\end{equation}
in which $t= k_BT/J$, $h=H/J$ and $d=D/J$. Since the free energy must be a global minimum in terms of the average magnetization $m$, we have 

\begin{equation}
m= \frac{\sinh[(m+h)/t]}{\textrm{e}^{d/t}+2\cosh[(m+h)/t]}+\frac{\sinh[(m-h)/t]}{\textrm{e}^{d/t}+2\cosh[(m-h)/t]}.
\label{eq:magnetization2}
\end{equation}

From equations (\ref{eq:lim_termo}) and (\ref{eq:magnetization2}) four phases are found at the ground state: (a) ferromagnetic phase, with $m\pm 1$, $q=\left\langle S^2\right\rangle$ and  $f=d-1/2$; (b) non-magnetic phase, in which $m=0$, $q=0$ and $f=0$; (c) paramagnetic phase, with $m=0$, $q=1$ and $f=d-h$; (d) ferromagnetic-non-magnetic phase, with $m=\pm 1/2$, $q=1/2$ and $f=(d-h-1/2)/2$. 

Although we have made use of the replica approach the resulting free energy density is exactly the same as that obtained by Kaufman and
Kanner, and so their $h-t$ phase diagrams are recovered. In accordance with recent works \cite{Santosetal,Salmon10},
we use the following notation to describe the structure of phase diagrams in this work: continuous transition lines are represented by continuous line; first-order transition lines are represented by dotted line; tricritical point is represented by a black circle; ordered critical point is represented by a black star; critical endpoint is represented by a black triangle and $A^5$ point is represented by a black square.

In order to determine critical lines and tricritical points, we make use of a Landau-like expansion:
\begin{eqnarray}
f(t, d; m) = A_0 + A_{2}m^2 + A_{4}m^4 + A_{6}m^6 + \cdots .
\label{eq:Landau}
\end{eqnarray}
\noindent
Using \cite{Kaufman} the coefficients $A_2$, $A_4$ and $A_6$ are given, respectively, by:
\begin{equation}
A_2 =\frac{J^2}{2!}\left[\frac{1}{J}+w^{2}u-w\right],
\label{eq:A2}
\end{equation}
\begin{equation}
A_4 =\frac{J^4}{4!}\left[6w^{4}u^2 -12w^{3}u+w^{2}(4u+3)-w\right],
\label{eq:A4}
\end{equation}
\begin{equation}
\begin{split}
A_6 = & \frac{J^6}{6!}[  120w^{6}u^{3}-360w^{5}u^2 +w^{4}(120u^2 +270u) \\ & -w^{3}(150u+30)+w^{2}(16u+15)-w],
\end{split}
\label{eq:A6}
\end{equation}
\noindent
where 
\begin{equation}
u=\tanh^{2}(JH),
\label{eq:u}
\end{equation}
\begin{equation}
w = 2\textrm{e}^{-JD}\cosh(JH)[1 + 2e^{-JD}\cosh(JH)]^{-1}.
\label{eq:w}
\end{equation}

The $A_4$ coefficient given by eq. (\ref{eq:A4}) differs slightly from Kaufman and Kanner \cite{Kaufman} results. In their paper, the $a_4$ coefficient presented in eq. (7) starts with $6w^{6}u^2$ inside the bracket. However, this difference does not make any change in the corresponding phase diagrams as long as we have analyzed. 

The critical lines are obtained by imposing $A_2=0$, while $ A_4>0$. Thus, the critical boundaries are determined by an implicit equation given by

\begin{eqnarray}
\frac{2\cosh(h/t)}{\textrm{e}^{d/t}+2\cosh(h/t)}-\left(\frac{2\sinh(h/t)}{\textrm{e}^{d/t}+2\cosh(h/t)}\right)^2=t ,
\label{eq:critical_linekk}
\end{eqnarray}
\noindent
provided that $A_4$ remains positive.

In order to obtain tricritical points we must impose the conditions

$$ A_2 = 0, ~~ A_4 = 0, \quad \mbox{while} ~~ A_6 > 0 . $$

If $A_6$ also vanishes along with the simultaneous vanishing of $A_2$ and $A_4$ we could have a higher order multicritical point.

In the following section some representative $d-t$ phase diagrams are presented, for several values of $h$. 

\section{Phase Diagrams}

The phase diagrams were obtained by numerically finding the global minimum of the free energy density given by Eq.~(\ref{eq:lim_termo}). According to the discussion above, the critical lines were analytically obtained in closed but implicit form, and thus they can be evaluated numerically.

\begin{figure}[!htbp]
	\centering
		\includegraphics[scale=0.4]{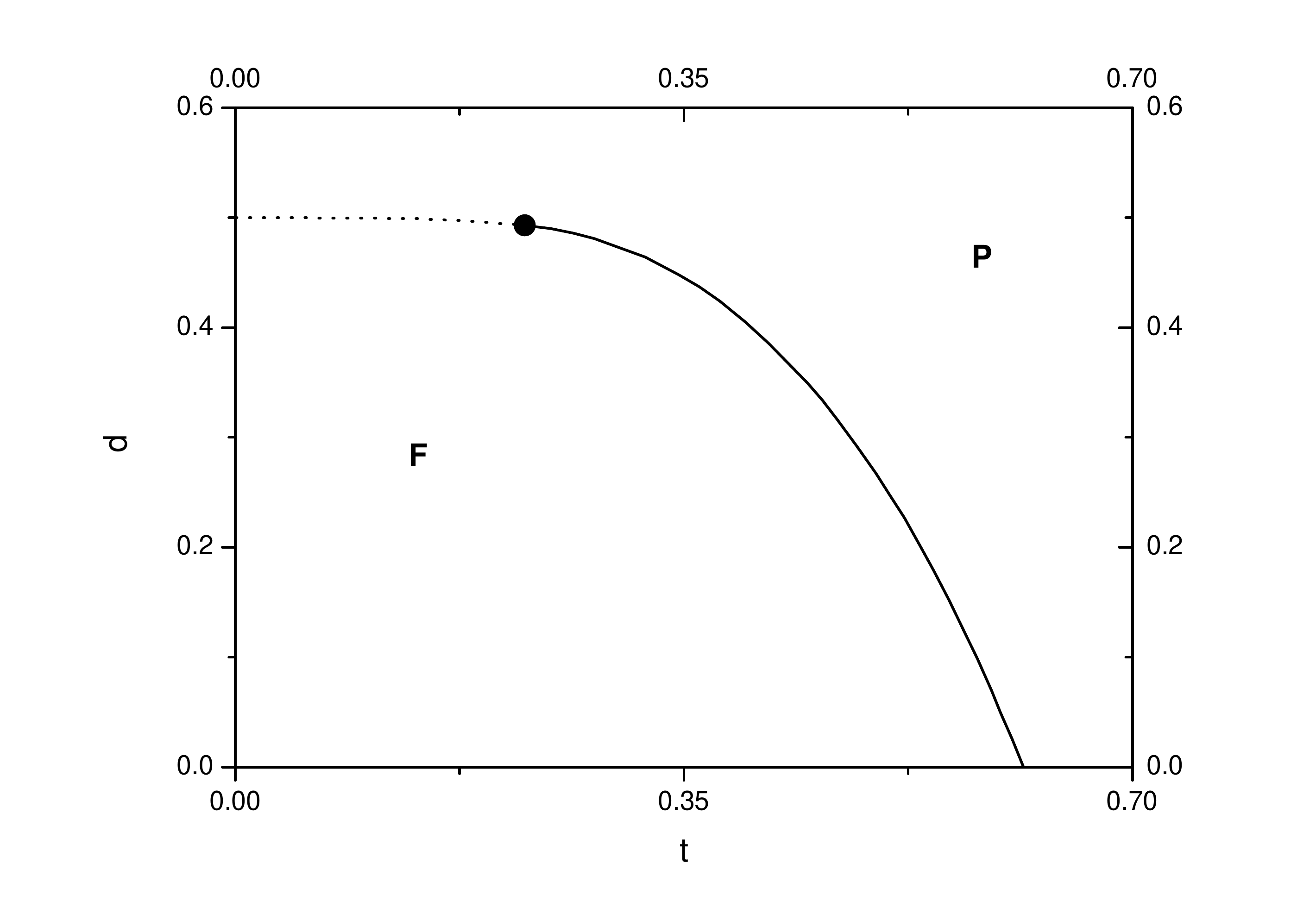}
	\caption{Phase diagram in plane $d-t$ for $h=0.25$ with a tricritical point (solid circle). The dotted line represents first-order transitions whereas the solid one is a continuous transition line.}
\label{fig:grafico_h025}
\end{figure}

\begin{figure}[!htbp]
	\centering
		\includegraphics[scale=0.4]{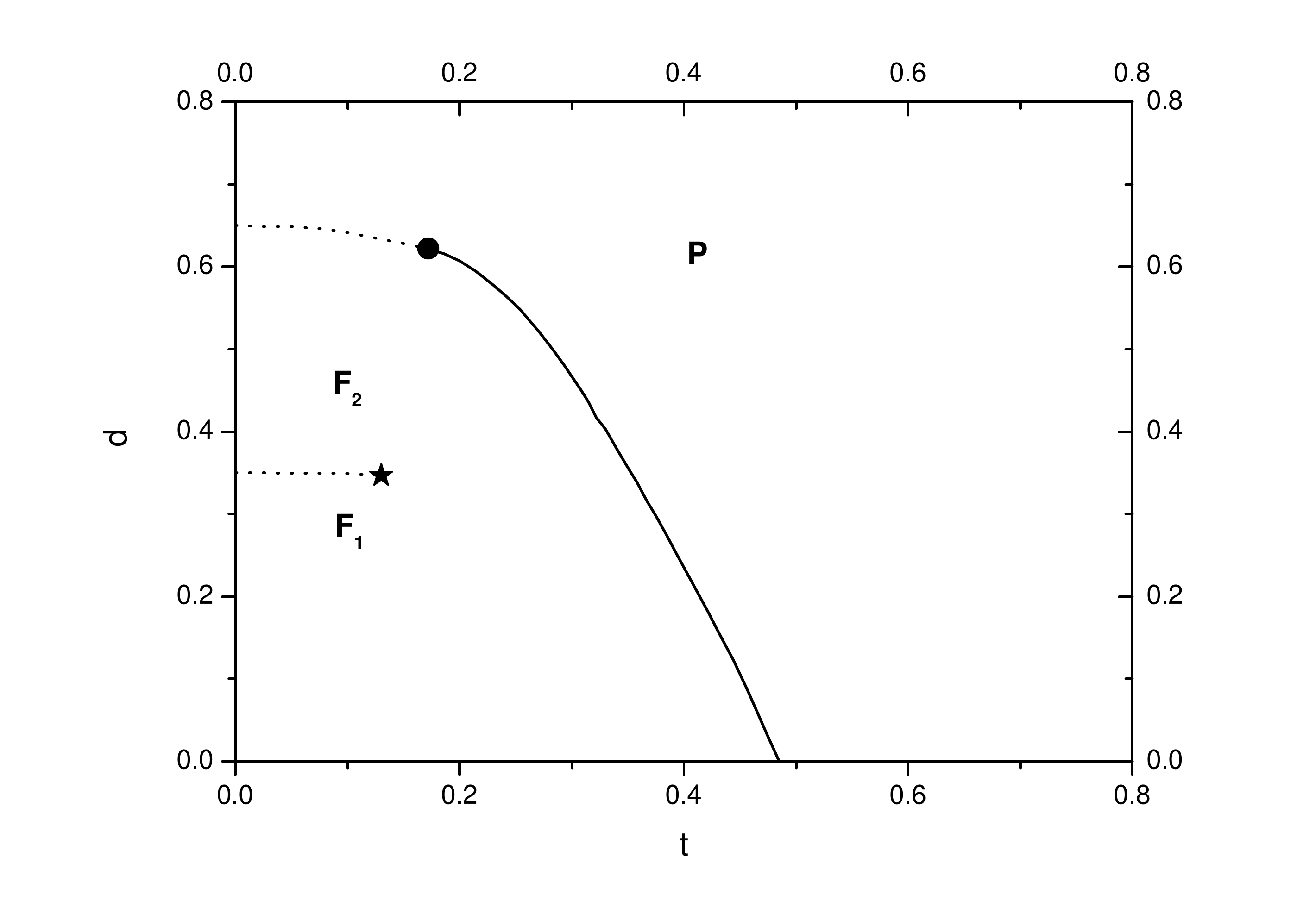}
	\caption{Phase diagram in plane $d-t$ for $h=0.4$ with a tricritical point (solid circle) and an ordered critical point (solid star). The dotted lines represent first-order transitions whereas the solid one is a continuous transition line.}
\label{fig:grafico_h04}
\end{figure}

\begin{figure}[!htbp]
  \centering
      \includegraphics[scale=0.4]{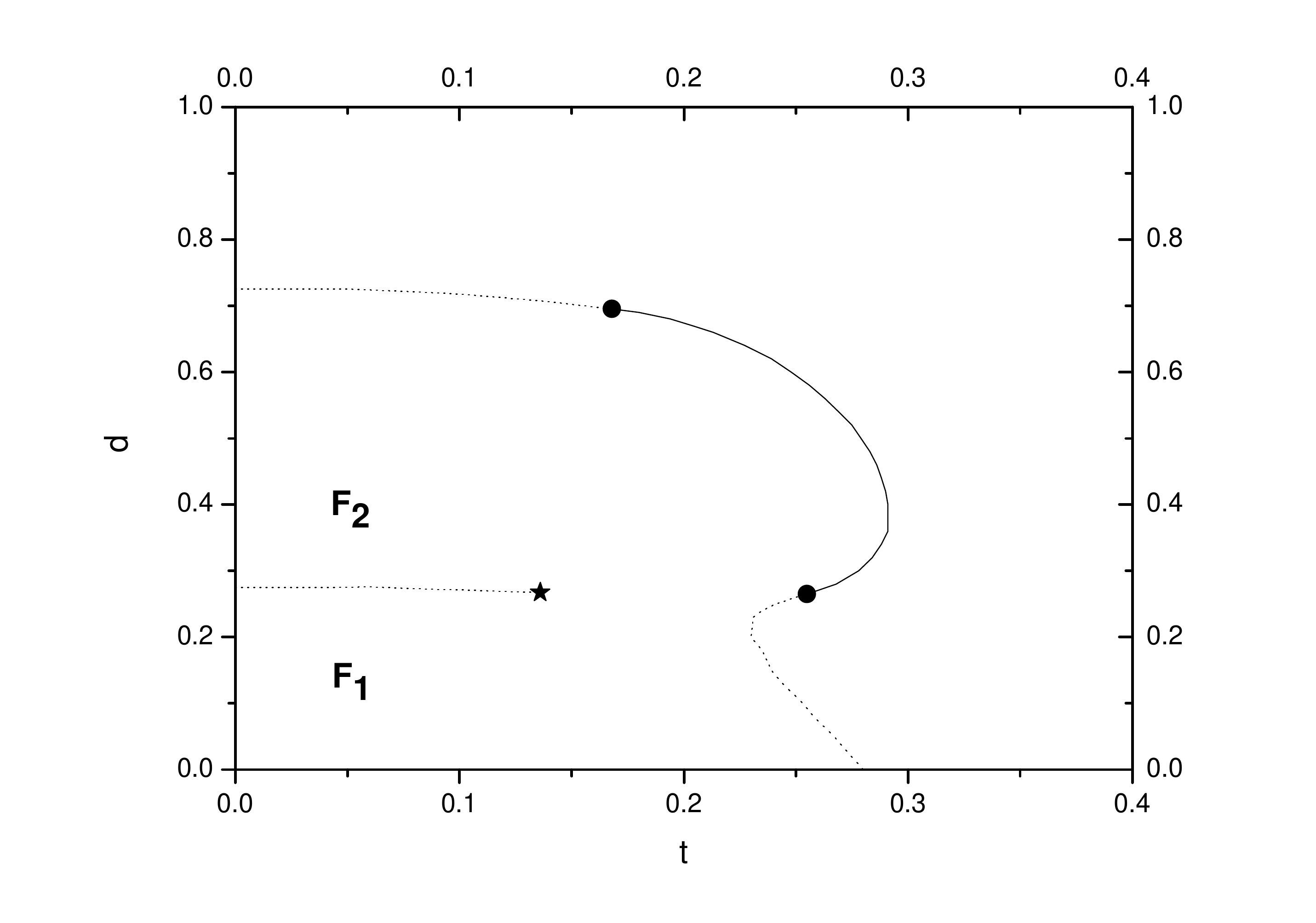}
      \caption{Phase diagram in plane $d-t$ for $h=0.475$, corresponding to Topology III, displaying two tricritical points (solid circle) and an ordered critical point (solid star). The dotted lines represent first-order transitions (coexistence of \textbf{$\mathbf{F_1}$-$\mathbf{F_2}$}, \textbf{$\mathbf{F_1}$-P} and \textbf{$\mathbf{F_2}$-P} phases) whereas the solid one is a continuous transition line.}
\label{fig:grafico_h0475}
\end{figure}

\begin{figure}[!htbp]
	\centering
		\includegraphics[scale=0.4]{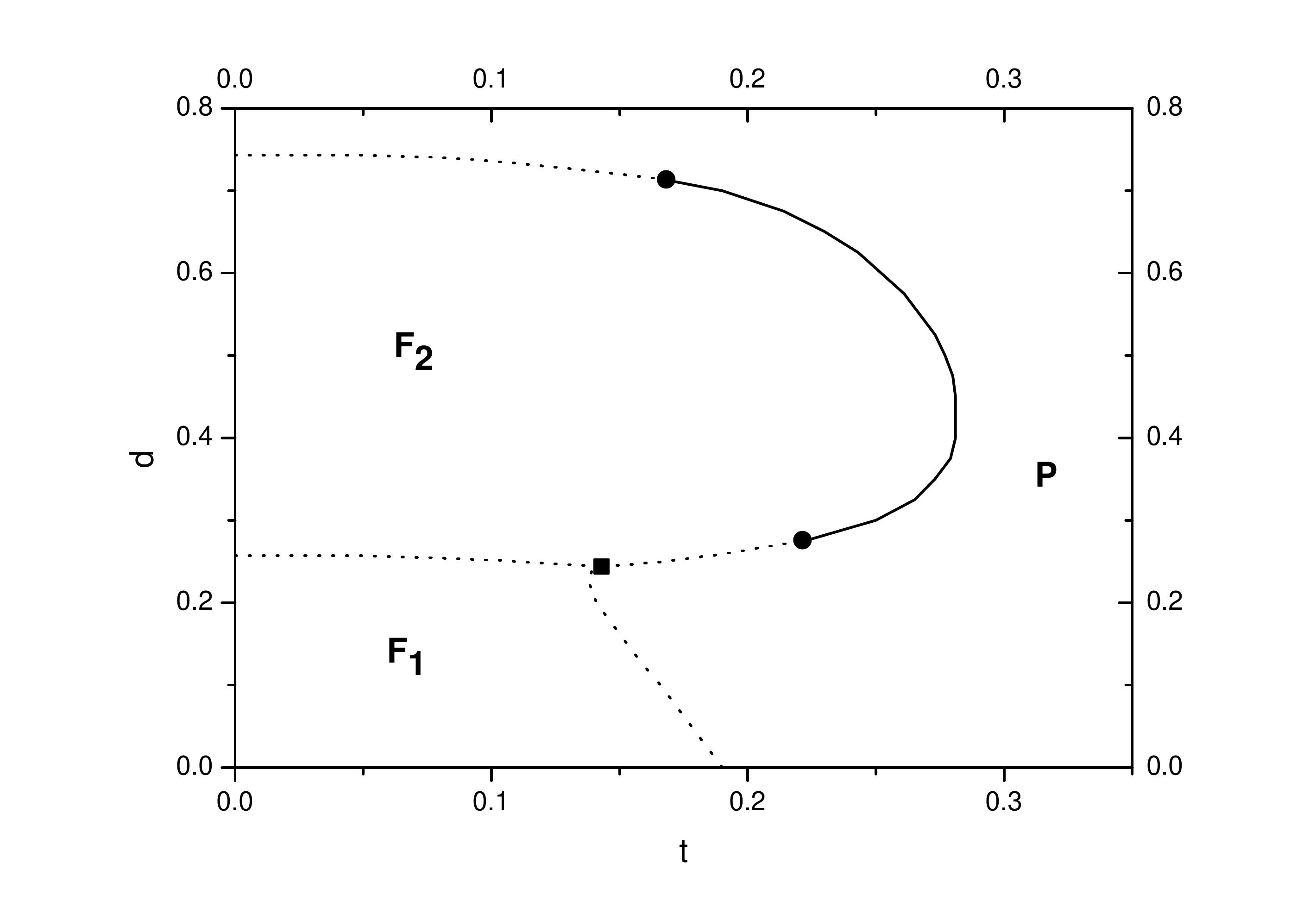}
	\caption{Phase diagram in plane $d-t$ for $h=0.493$ with two tricritical points (solid circles) and a $A^5$ point (solid square),  as 5 phases coexist: 1 paramagnetic, 2 ferromagnetic ($\mathbf{F_1}$) and 2 ferromagnetic ($\mathbf{F_2}$). The dotted lines represent first-order transitions whereas the solid one is a continuous transition line.}
\label{fig:grafico_h0493}
\end{figure}

\begin{figure}[!htbp]
	\centering
		\includegraphics[scale=0.4]{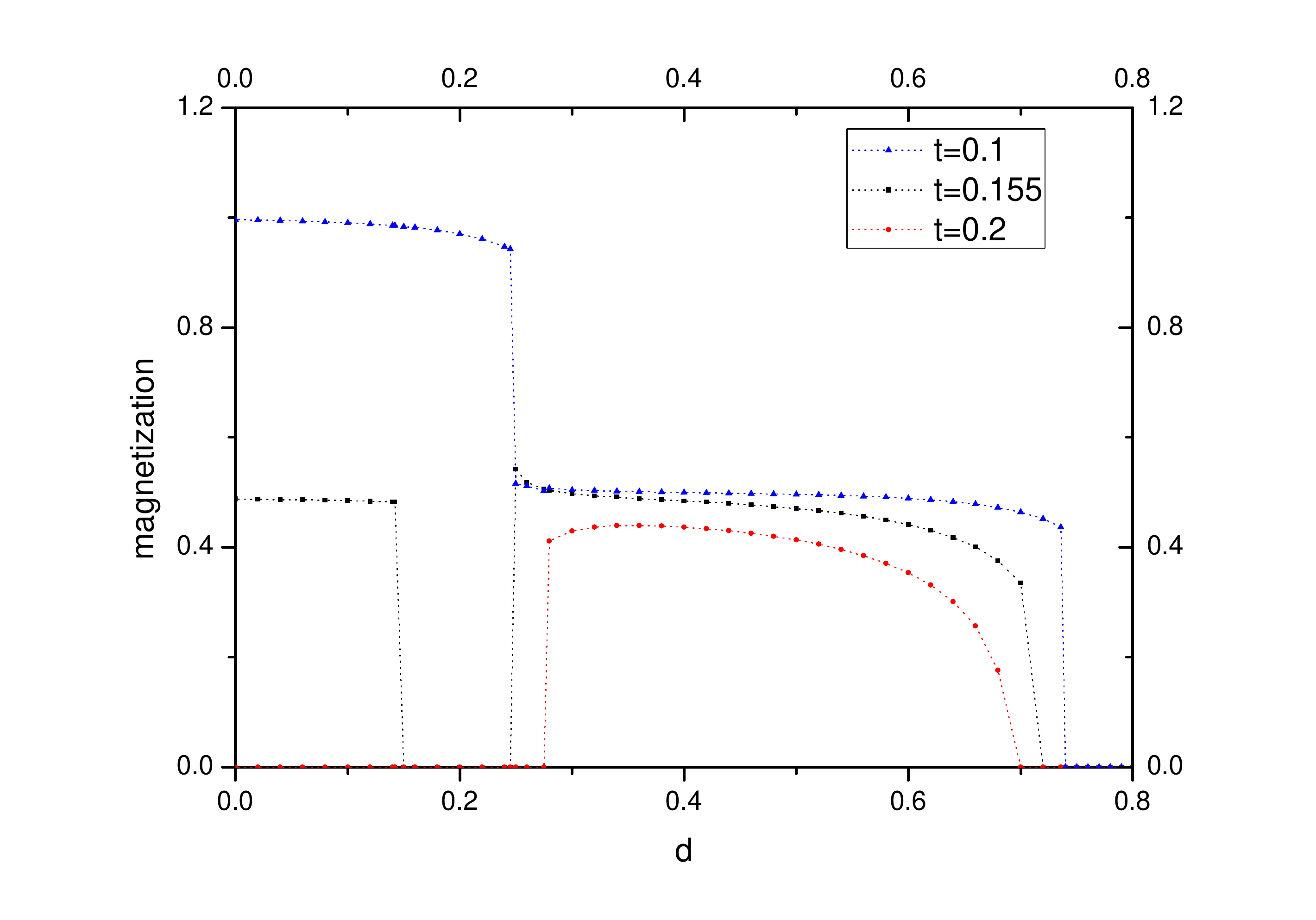}
	\caption{Magnetization profiles for h=0.493 as a function of d for some values of temperature:  t=0.10 (blue curve), t=0.155 (black curve) and t= 0.20 (red curve).}
\label{fig:magcurve_h0493}
\end{figure}

\begin{figure}[!htbp]
	\centering
		\includegraphics[scale=0.4]{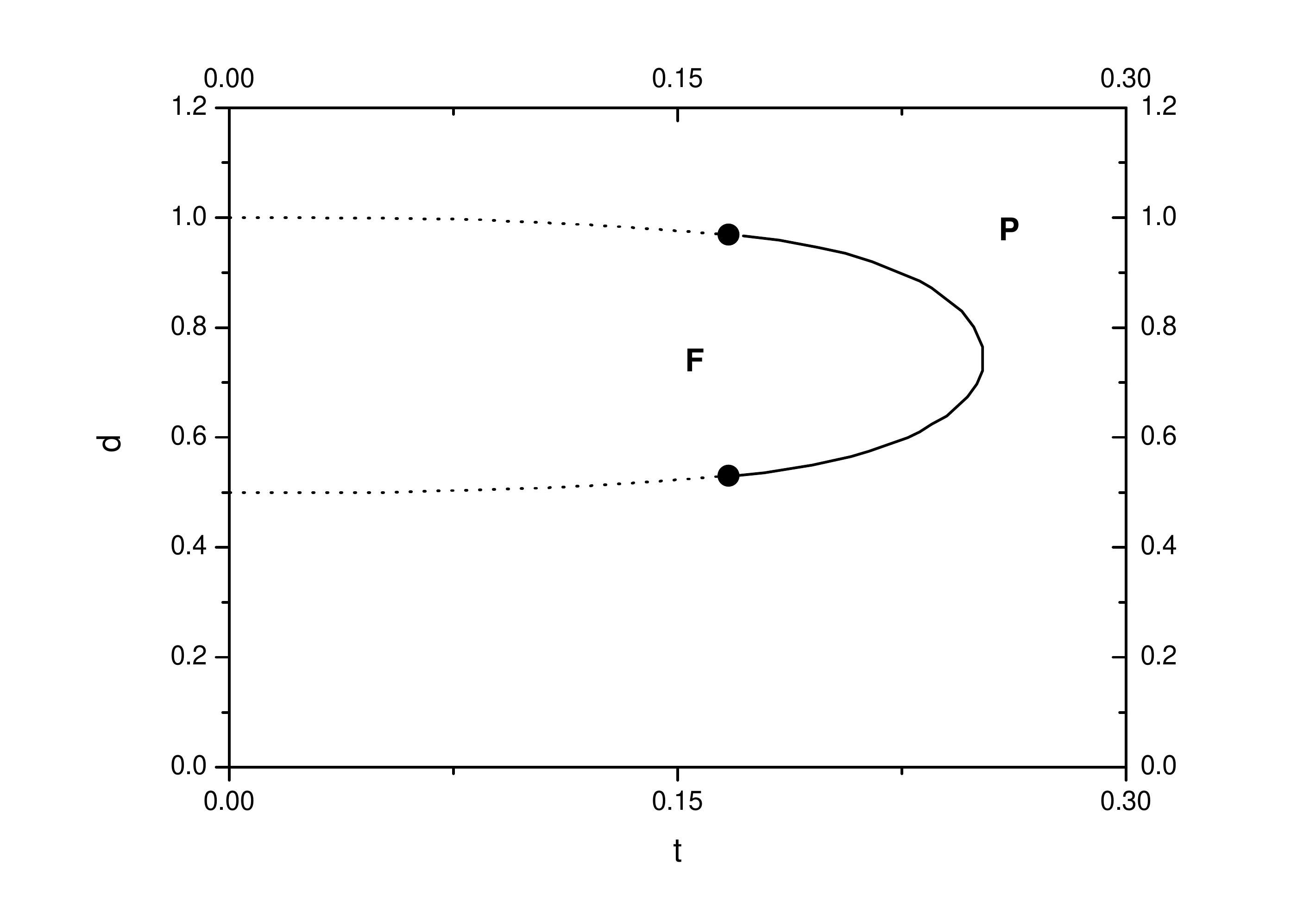}
	\caption{Phase diagram in plane $d-t$ for $h=0.75$ displaying two tricritical points (solid circles), connected to each other by a critical (solid) line. The dotted lines represent first-order transition lines.}
\label{fig:grafico_h075}
\end{figure}

In Figs.~(\ref{fig:grafico_h025}- \ref{fig:grafico_h075}) we present some representative $d-t$ phase diagrams for distinct values of $h$. These phase diagrams represent the possible topologies which arise as parameter $h$ is varied. For $h=0.25$ we have a phase diagram belonging to Topology I, as shown Fig. \ref{fig:grafico_h025}. This sort of phase diagram is similar to the case of the Blume-Capel model in zero field: the ferromagnetic phase is separated from  paramagnetic one by a critical line for sufficiently high temperature, whereas for low temperatures these two phase may coexist along a first-order transition line. Those two lines merge at a tricritical point. 

A typical case of Topology II occurs for $h=0.4$ and is shown in Fig. \ref{fig:grafico_h04}. For low temperatures, besides the coexistence line between the paramagnetic and the ferromagnetic phases, we found another first-order transition line separating two ordered phases, $\mathbf{F_1}$ and $\mathbf{F_2}$. That line ends at an ordered critical point.

Fig. \ref{fig:grafico_h0475} displays an example of Topology III phase diagrams. For this kind of topology we found three first-order transition lines, one of such lines inside the ordered phase region separating the ferromagnetic $\mathbf{F_1}$ and $\mathbf{F_2}$ phases. Each of the other two first-order transition lines separating ferro and paramagnetic phases ends at a tricritical point. These two tricritical points are connected to each other by a ferromagnetic to paramagnetic critical line. Inside the ordered phase the first-order transition line ends at an ordered critical point.  

A typical phase diagram representing Topology IV is shown in Fig. \ref{fig:grafico_h0493} for $h=0.493$. For this case we have two tricritical points, one critical line and four coexistence lines. Three of those coexistence lines merge together at a $A^5$ point (in fact there are two $\mathbf{F_1}$ phases corresponding to opposite magnetization and the same applies to $\mathbf{F_2}$ phases). For $h=0.493$, Fig. \ref{fig:magcurve_h0493} shows the interesting behavior of magnetization as a function of external field $d$ for some relevant values of temperature. For $t=0.10$ (blue curve) there are two first-order transitions: \textbf{$\mathbf{F_1}-\mathbf{F_2}$}, then \textbf{$\mathbf{F_2}$ - P}. For $t=0.155$ (black curve) there are three first-order transitions: \textbf{$\mathbf{F_1}$ - P}, then \textbf{P -  $\mathbf{F_2}$}, and finally \textbf{$\mathbf{F_2}$ - P}. For t= 0.20 (red curve) there is a first-order transition \textbf{P - $\mathbf{F_2}$} followed by a continuous transition \textbf{$\mathbf{F_2}$ - P}. Finally, for $h>0.5$  we have phase diagrams belonging to Topology V, as is displayed in Fig. \ref{fig:grafico_h075} for $h=0.75$. In that case we obtain two first-order transition lines, each of them separating the ordered, ferromagnetic, phase from the paramagnetic one. Again, these coexistence lines end at two distinct tricritical points which are connected by a continuous transition line representing the critical bordering between the paramagnetic and ferromagnetic phases.

 \newpage
\section{Conclusions}

In this paper we have revisited the mean-field analysis previously considered by Kaufman and Kanner for the Blume-Capel model under a bimodal random field \cite{Kaufman}. We have re-obtained their results and confirmed the structure of their random magnetic field versus temperature phase diagrams. In recent studies much attention has been given to crystal field anisotropy versus temperature phase diagrams for random spin-1 models, especially in connection to reentrant effects in the vicinity of tricritical points. Thus, we decided to complement the work initiated by Kaufman and Kanner by investigating the crystal field versus temperature phase diagrams for given values of the random magnetic field. As has been reported in recent works, the obtained phase diagrams were classified within five distinct topologies: (i) Topology I, displaying both a continuous and a first-order transition lines, separating the paramagnetic and ferromagnetic phases, with a tricritical point at the merging of these two lines. (ii) Topology II, in which the paramagnetic phase is separated from the ordered ferromagnetic phase as in the case of topology I. However, we also find a first-order transition line inside the ordered phase separating two distinct ferromagnetic phases $\mathbf{F_1}$ and $\mathbf{F_2}$. This coexistence line ends at an ordered critical point. (iii) Topology III, displaying a richer phase diagram in which the paramagnetic phase is separated from the ordered one by both first-order and continuous transition lines with the appearance of two tricritical points. Within the ordered phase we also find $\mathbf{F_1}$ and $\mathbf{F_2}$ ferromagnetic phases coexisting along line which ends at an ordered critical point. (iv) Topology IV, which displays a richer behavior with the presence of first-order and continuous transition lines separating the ordered phase from the paramagnetic ones and the presence of two tricritical points. We also find three distinct first-order transition lines meeting at an usual $A^5$ point and obeying to the $180^{o}$ rule. (v) Finally, we have Topology V, in which the ordered ferromagnetic phase is bounded by first-order lines at low temperatures and by a continuous transition line at intermediate temperatures. There is also two tricritical points at the meeting of first-order and continuous transition lines. As in the case of the phase diagrams obtained by Kaufman and Kanner \cite{Kaufman}, our findings indicate that no reentrant behavior occurs in crystal field versus temperature phase diagrams.

\section*{Acknowledgement}
The authors would like to thank Coordena\c c\~ao de  Aperfei\c coamento de Pessoal de N\'ivel Superior (CAPES) for partial financial support. 
%
\newpage
%

%
%

\begin{thebibliography}{99}
\bibitem{Kaufman} M. Kaufman and M. Kanner, Phys. Rev. B \textbf{42}, 2378 (1990). 
\bibitem{DeDominicis_Giardina} C. De Dominicis and I. Giardina, \textit{Random Fields and Spin Glasses} (Cambridge University Press, Cambridge, 2006).
\bibitem{Gulpinaretal} G. Gulpinar, R. Erdem and M. Agartioglu,  J. Magn. Magn. Mater \textbf{439}, 44 (2017).
\bibitem{Acharyya} M. Acharyya and A. Halder, J. Magn. Magn. Mater \textbf{426}, 53 (2017).
\bibitem{daSilva} W.P. da Silva, P.H.Z. de Arruda, T.M. Tunes, M. Godoy and A.S. de Arruda, J. Magn. Magn. Mater. \textbf{422}, 367 (2017).
\bibitem{Sumedha} Sumedha and N. K. Jana, J. Phys. A: Math. Theor.\textbf{50}, 015003 (2017).
\bibitem{Carvalho} D.C. Carvalho and J.A. Plascak, Physica A \textbf{432}, 240 (2015). 
\bibitem{Costabile} E. Costabile, J. R. Viana, J. R. Souza and A. S. Arruda, Sol. Stat. Com. \textbf{212}, 30 (2015). 
\bibitem{Santosetal} P. V. Santos, F.  A. da Costa and J. M. de Ara\'ujo, Phys. Lett. A \textbf{379}, 1397 (2015).
\bibitem{Nishimori} H. Nishimori, \textit{Statistical Physics of Spin Glasses and Information Processing} (Oxford University Press, Oxford, 2001).
\bibitem{Kaufman2} M. Kaufman, P.E. Klunzinger and A. Khurana, Phys. Rev. B \textbf{34}, 4766 (1986). 
\bibitem{Blume} M. Blume, Phys. Rev. \textbf{141}, 517 (1966). 
\bibitem{Capel} H. W. Capel, Physica \textbf{32}, 966 (1966). 
\bibitem{BlumeEmeryGriffiths} M. Blume, V.J. Emery and R.B. Griffiths, Phys. Rev. A \textbf{4}, 1071 (1971).
\bibitem{Danielle} D.A. Dias and J.A. Plascak, Phys. Lett. A \textbf{375}, 2089 (2011).
\bibitem{Lara09} D. P. Lara, G. A. P. Alcazar, L. E. Zamora and J. A. Plascak, Phys. Rev. B \textbf{80}, 014427 (2009).
\bibitem{Blatter} A. Blatter and M. von Allmen, Phys. Rev. Lett. \textbf{54}, 2103 (1985).
\bibitem{Cowlam} W. Sinkler, C. Michaelsen, R. Bormann, D. Spilsbury and N. Cowlam, Phys. Rev. B \textbf{55}, 2874 (1997).
\bibitem{Mazzitello} K. I. Mazzitello, J. Candia and E. V. Albano, Phys. Rev. E \textbf{91}, 042118 (2015).
\bibitem{Goldman} A. M. Goldman, Phys. Rev. Lett. \textbf{30}, 1038 (1973).
\bibitem{Giordano} N. Giordano and W. P. Wolf, Phys. Rev. Lett. \textbf{39}, 342 (1977).
\bibitem{Johari} G. P. Johari, Physiol. Chem. Phys. \textbf{3}, 2483 (2001).
\bibitem{Taylor} P. E. Cladis, R. K. Bogardus, W. B. Daniels and G. N. Taylor, Phys. Rev. Lett. \textbf{39}, 720 (1977).
\bibitem{Antenucci} F. Antenucci, A. Crisanti and L. Leuzzi, J. Stat. Phys. \textbf{155}, 909 (2014).
\bibitem{Mahan} G. D. Mahan and S. M. Girvin, Phys. Rev. B 17, 44119 (1978).
\bibitem{Clusel} M. Clusel, J. Y. Fortin and V. N. Plechko, J. Phys. A: Math. Theor. \textbf{41}, 405004 (2008).
\bibitem{Berker} A. N. Berker, J. Appl. Phys. \textbf{70}, 5941 (1991). 
\bibitem{Hui} K. Hui and A. N. Berker, Phy. Rev. Lett. \textbf{61}, 2507 (1989). 
\bibitem{Imry} Y. Imry and M. Wortis, Phy. Rev. B \textbf{19}, 3580 (1979).
\bibitem{Lara12} D.P. Lara, Rev. Mex. Fis. \textbf{58}, 203 (2012). 
\bibitem{Branco97} N. S. Branco and B. M. Boechat, Phys. Rev. B \textbf{56}, 11673 (1997). 
\bibitem{Benyoussef87} A. Benyoussef, T. Biaz, M. Saber and M. Touzani, J. Phys. C: Solid State Phy.\textbf{20}, 5349 (1987). 
\bibitem{Carneiro89} C. E. I. Carneiro, V. B. Henriques and S. R. Salinas, J. Phys.: Condens. Matter \textbf{1}, 3687 (1989). 
\bibitem{Salmon10} O.D.R. Salmon and J.R. Tapia, J. Phys. A: Math. Theor. \textbf{43}, 125003 (2010). 
\bibitem{Boccara} N. Boccara, A. Elkenz and M. Saber, J. Phys., Condens. Matter \textbf{1}, 5721 (1989). 
\bibitem{Maritan} A. Maritan, M. Cieplak, M.R. Swift, F. Toigo and J.R. Banavar, Phys. Rev. Lett. \textbf{69}, 221 (1992) 
\bibitem{Kaneyoshi} T. Kaneyoshi and J. Mielnicki, J. Phys. Condens. Matter \textbf{2}, 8773 (1990). 
\bibitem{Buzano} C. Buzano, A. Maritan and A. Pelizzola, J. Phys. Condens. Matter \textbf{6}, 327 (1994). 
\bibitem{Borelli} M.S.E. Borelli, C.E.I. Carneiro, Physica A \textbf{230}, 249 (1996). 
\bibitem{Albayrak11} E. Albayrak, Physica A \textbf{390}, 1529 (2011). 
\bibitem{Kaya} T. Kaya and M. Yavuz, Int. J. Mod. Phys. B \textbf{24}, 5771 (2010).
\bibitem{Zhang} Y.F. Zhang  and  S.L. Yan, Phys. Lett. A \textbf{372}, 2696 (2008).
\bibitem{daCosta} F. A. da Costa, Phys. Rev. B \textbf{82}, 052402 (2010). 
\bibitem{Ferrari} U. Ferrari and L. Leuzzi, J. Stat. Mech., P12005 (2012).
\bibitem{Binder and Young} K. Binder and A. P. Young, Rev. Mod. Phys. \textbf{58}, 801 (1986).
\end{thebibliography}
\end{document}